\documentclass[12pt]{article}
\newcommand{\beq}{\begin{equation}\displaystyle}
\newcommand{\eeq}{\end{equation}}
\newcommand{\bit}{\begin{itemize}}
\newcommand{\eit}{\end{itemize}}
\newcommand{\ben}{\begin{enumerate}}
\newcommand{\een}{\end{enumerate}}
\newcommand{\bc}{\begin{center}}
\newcommand{\ec}{\end{center}}

\begin{document} 
\baselineskip 19pt

\bc \large \bf Velocity-selective resonance dips in the probe absorption spectra of
Rb D$_{2}$ transitions induced by a pump laser.
\ec  \vskip 2.5cm

  \vskip 1.5cm

\noindent {\bf PACS number(s): 32.60 + i,   32.80. Bx,   33.80. Be}  \vskip 3.5cm

\begin{center} \bf\normalsize S. Chakrabarti, A. Pradhan, A. Bandyopadhyay,
A. Ray, B. Ray, N. Kar \footnote[2]{Permanent address: Department of Physics,
North Bengal University, Siliguri - 734 430 , India.}
\\ and \\ P. N. Ghosh\,\footnote[1]{Corresponding author.} \\ \,Department of
Physics, University of Calcutta \\ 92, A. P. C. Road, Calcutta --
700 009, \\ I N D I A \\ (Phone : +91 33 2350 8386, Fax : +91 33
2351 9755, \\ E--mail : speclab@cubmb.ernet.in) 
\end{center}

\newpage\begin{abstract}\baselineskip 22pt
We report experimental observation of velocity-selective resonances in the
Doppler-broadened probe absorption spectra of $^{85}$Rb and $^{87}$Rb D$_{2}$
transitions in the presence of a strong copropagating pump  laser locked to
a frequency within the Doppler profile of the transition. The set of three dips
having the separation of allowed hyperfine transitions can be moved along the
Doppler profile by tuning the pump laser frequency indicating a resonance
between the pump laser frequency and the velocity shifted probe laser
frequency.

\end{abstract}

\newpage \noindent \vskip 4.5mm

Manipulation of atomic response to a probe radiation field by using an
intense pump laser has attracted considerable attention in recent years.
Typical atomic systems considered for the purpose are three level atoms
in the ladder or $\Lambda$ configuriation. The absorption profile of
an atomic transition when the upper level is coupled coherently to a third
level by a strong laser field exhibits reduction of absorption [1] leading
to electromagnetically induced transperency(EIT). Since the pioneering
work on quantum interference leading to zero absorption at the line centre
by Harris and co-workers [2] and Agarwal and co-workers [3,4] a large number
of theoretical and experimental observations on EIT were reported [5-9]. Most of
the experimental work reported observation of EIT in a $\Lambda$ type system
in Rb. These work involved two lower hyperfine levels with a common upper
hyperfine level in a D transition. The hyperfine split components of D$_{1}$
or D$_{2}$ transitions in alkali atoms are embedded in a Doppler broadened
Gaussian background that makes it impossible to measure the splitting by
using single resonance spectroscopy [10]. In order to recover the hyperfine
components from the Doppler background standing wave Lamb dip spectroscopic
method is  used [10,11]. This method uses two counter propagating laser beams
originating from the same source and crossing each other within the gas cell.
If one of the beams is stronger than the other then it saturates the transition
and creates a hole in the lower level Gaussian population distribution.
At resonance the probe absorption shows a dip in the absorption spectra
because of the population depletion of the lower level. If there are two
closely spaced upper levels, in addition to the two dips at resonance
frequencies $\Omega_{1}$ and $\Omega_{2}$ an additional "crossover"
resonance is observed at a frequency intermediate between the two frequencies
i.e. at $\frac{(\Omega_{1} + \Omega_{2})}{2}$ [12,13]. Such a resonance
occurs when the frequency tuned probe absorption of one beam meets the hole
created by the other beam. Hence, in such cases three resonance dips are
observed all of which may have similar strength [13]. In the case of hyperfine
spectroscopy the atomic energy levels commonly involved are a single ground state
and three closely spaced upper level, because of the selection rule
$\Delta{F}$ = 0, $\pm$ 1. If the lower state hyperfine splitting is much
larger the transitions involving different lower levels do not overlap and
can be observed as seperate transitions, commonly used in $\Lambda$ type
systems. For the three hyperfine transitions Lamb dip spectroscopy reveals
three Lamb dips and three crossover resonance dips. The Lamb dips for $^{85}$Rb
(F = 2 $\rightarrow$ F' and F = 3 $\rightarrow$ F') and for $^{87}$Rb
(F = 1 $\rightarrow$ F' and F = 2 $\rightarrow$ F') have been studied
in detail [14], mainly because of the importance in laser cooling and
Bose-Einstein Condensation with Rb. Usually the hyperfine transition F=2 $\rightarrow$
F'=3 in $^{87}$Rb is frequency locked for using it as cooling transition. However,
all the six components may not be fully resolved because of the close spacing of
the six frequencies. Recent theoretical simulation [15] shows the difficulty of
isolating the components in Lamb dip spectroscopy.

        In the traditional methods of double resonance spectroscopy in a three level
system with one common level copropagating pump and probe beams are used
to observe induced emission power of the probe [16,17] in the case of
homogeneously broadened microwave-microwave transitions. The emission
line shape and a splitting of the absorption peak may be observed depending on the
power of the saturating beam [17]. In the case of inhomogeneously broadened
transitions like D$_{1}$ and D$_{2}$ transitions in the alkali atoms similar
copropagating beams from two laser sources may be used. Very recently such a
pump-probe spectroscopic method using copropagating laser radiations has been
reported to observe additional non-linear resonances of the Sodium D$_{1}$ line [18].
They used a four level system with two lower hyperfine levels and two upper
hyperfine levels. The system is a combination of double $\Lambda$ and
two V-type transitions. In addition to the EIT of the $\Lambda$-type
resonance they reported difference frequency crossing in the V-type
configuration. We report a pump-probe experiment on Rb D$_{2}$ lines using copropagating
laser beams where the pump beam(or the control laser) is locked at
a certain frequency within the Doppler broadened background, the
probe frequency is tuned and the probe power is monitored. At all
the lock points of the pump beam we observe very clearly a set of
three dips corresponding to the hyperfine components. The set of dips can be
shifted by tuning the pump frequency thus exhibiting velocity selective resonances(VSR).
Such a set of drifting VSR dips are reported for the first time.

The pump-probe experiment carried out in this work uses an atomic rubidium vapour cell having
a length of 5cm and window of diameter 2.5 cm (Fig.\,1). The vapour cell is sealed with a
pressure of one micro-Torr and there is no buffer gas. The pump frequency is produced
by an external cavity diode laser(ECDL), TEC 100 from Sacher Lasertechnik. The weak
probe frequency is obtained from another ECDL, Velocity 6312 from New Focus. The pump beam
is split into two components using a 70:30 beam splitter. The transmitted beam is sent 
through another cell (cell 2) of the same configuration for saturation absorption spectroscopy (SAS) or Lamb dip measurements.
The Lamb dip signal consisting of the hyperfine  and  crossover resonance transitions is 
fed to the input of the PID lock circuit for frequency locking purpose.
The pump frequency is locked at different frequencies around the $^{85}$Rb 5$^{2}S_{\frac{1}{2}}$(F = 3)
$\rightarrow$ 5$^{2}P_{\frac{3}{2}}$(F$^{'}$ = 2, 3, 4) transitions (Fig.\,2) and their crossover resonance
peaks appearing in the Doppler broadened background by using a home built PID lock
circuit that has a long term frequency stability of 1 MHz. The reflected pump beam is sent through cell 1 .
The pump and probe fields in cell 1 are copropagating and colinear.
A Coherent wavemeter is used to monitor the wavelengths of the pump and probe trasitions. 
The typical pump and probe intensities used in the sample cell are 22 mW/cm$^{2}$ and 1.27 mW/cm$^{2}$ respectively.
The powers are measured by a Melles Griot power meter. The lock point is varied using the offset of the lock circuit from the
lower frequency to the higher frequency region. For each lock frequency of the pump,
the probe scans a frequency range of nearly 1 GHz around the $^{85}$Rb D$_{2}$
(F = $\rightarrow$ F$^{'}$) transitions. The pump beam is sent through an optical isolator
to avoid feedback of laser power. Fig.\,3a is the probe transmission in the absence
of pump field and this may be used as the reference spectrum. Figs.\,3b-3f show the
probe transmission in presence of the pump field with different lock frequencies as
shown. The set of three dips in the Doppler broadened background of probe absorption
shift from the lower frequency to the higher frequency region in harmony with the
increasing lock frequency of the pump laser. The third dip has the highest intensity 
and it always coincides with the pump frequency. The separations between the dips match with
the known hyperfine splittings ($\Omega_{2}$ - $\Omega_{1}$ = 63.43 and 
$\Omega_{3}$ - $\Omega_{2}$ = 120.91 MHz) of the excited state hyperfine lavels
5$^{2}P_{\frac{3}{2}}$ (F$^{'}$ = 2, 3, 4) of $^{85}$Rb [14]. The relative strengths of the dips
also change with shift of the lock point of the pump field. The set of three
dips representing emission or reduction of absorption, and their shifts are reproducible with
all pump powers larger than the probe power. By changing the pump power the peak height
and width of the dips can be changed. The collision broadening is negligible at the pressure used.
So the dips are power broadened Lorentzians.

The same experiment is repeated for the other D$_2$ transitions of  $^{85}$Rb and  $^{87}$Rb.
In all cases the hyperfine triplets moving with the pump laser frequency are observed.
However they are much weaker than those in Fig.\,3. Three well separated peaks in $^{87}$Rb
(F = 2 $\rightarrow$ F$^{'}$) are shown in Fig.\,4. Since one of the dips is weak vertical arrows are 
used to mark them.The separation being very large, with
the movement of the pump frequency, all the dips cannot be seen, as one of them may
fall beyond the Gaussian curve (Fig.\,4f). In fact this dip is very weak, but distinctly visible in other cases (Fig.\,4b-4e).
The observed separations of the triplet are $\Omega_{2}$ - $\Omega_{1}$
= 157.09 and $\Omega_{3}$ - $\Omega_{2}$ = 267.17 MHz. This agrees with the known hyperfine
splitting values [14]. In all the spectra the intermediate dip is the strongest and it coincides 
with the pump frequency.     
    
    When the pump laser is operating at any frequency within the Doppler broadened 
envelope, it can cause non-resonant excitations for all the transitions $\Omega_{1}$,
$\Omega_{2}$, and $\Omega_{3}$. Considering one-dimentional motion of atoms, this will
happen for the velocities $v_{i}$ of atoms given by 
$\Omega_{i}$ = $\omega_{pu}$ $(1- \frac{v_i}{c})$, i=1,2,3. These excitations
will cause population holes for the three velocity groups $v_{1}$, $v_{2}$  
and $v_{3}$. The atoms will meet the probe laser at shifted frequencies 
$\omega_{pr}$ $(1- \frac{v_i}{c})$ for these velocities. So the tuned probe laser
absorption will exhibit dips at $\omega_{pr}$ = $\omega_{pu}$,
$\omega_{pr}$ - $\omega_{pu}$ = $\pm$ $|(\Omega_{3}-\Omega_{1})|$,
$\pm$ $|(\Omega_{3}-\Omega_{2})|$, $\pm$ $|(\Omega_{2}-\Omega_{1})|$.
In principle there may be seven dips in the absorption curve.
    
     However, the dips will not all be of the same magnitude as the population 
depletion for the three velocity groups will not be the same. The population 
depletion will depend on two factors - (i) the number of atoms with velocity 
$v_{i}$:- $N_{i}$, which will be obtained from the Maxwell-Boltzmann velocity
distribution and (ii) the fraction $f_{i}$ of the atoms excited to the higher
state by the transition of frequency $\Omega_{i}$,$f_{i}$ will be determined 
by the transition probability of the particular case. It will also depend on the 
power of the pump laser but that is kept constant during the experiment. So if
$n_{i}$ is the depth of holes for the transition $\Omega_{i}$, we may write 
$n_{i}$ = $f_{i}$$N_{i}$. Simple numerical estimates show that for $\omega_{pu}$
in the Doppler broadened region, the velocities $v_{i}$ for the non-resonant 
excitations are such that the variations in $N_{i}$ are not very large if the 
hyperfine splitting is small. However, the relative transition probabilities for the 
three hyperfine transitions may be quite different. If one particular $f_{i}$
is much larger compared to the other two, then $n_{i}$, the depth of holes for that
particular transition will be much larger compared to the other two. For the
$^{85}$Rb D$_{2}$ transitions, it is known that $\Omega_{3}$ has the strongest
transition probability [19]; in that case $f_{3}$ is much greater than $f_{1}$ or $f_{2}$ and so the 
population depletion for the $v_{3}$ group will be the largest. One can then see that
the strongest dips will occur whenever $\omega_{pr}$ = $\omega_{pu}$ or
$\omega_{pr}$ = $\omega_{pu}$ - $|(\Omega_{3}-\Omega_{2})|$ or
$\omega_{pr}$ = $\omega_{pu}$ - $|(\Omega_{3}-\Omega_{1})|$ as long as $\omega_{pu}$
is within the Doppler-broadened region. The experimental observation agrees with this 
(Fig.\,3). In principle, there could be four other peaks, but they are too weak to be
observed by our experimental setup.

    The data for the $^{87}$Rb transitions can also be explained by similar
arguments. However, in this case the separations between the hyperfine transitions are
much larger. Hence $N_{i}$ may have wide variations. In this case we have to calculate
the complete absorption coefficient by considering the absorption cross section 
and the lower level population depletion caused by the non-resonant pump in order to 
determine the relative strength of the dips.                    

The experimental results obtained by us are similar to the difference frequency
crossing proposed by Wong et al.[18]. But we do not observe symmetrically located dips on
two sides of the pump laser frequency. Such dips will be very small and have negligible
intensity. The two symmetrical dips reported for the single V -type
system by Wong et al.[18] also had large difference in intensities. In our case of a double V - system
(single lower level and three upper levels) considering one dimensional motion of atoms we
predict seven dips. But the intensities of such dips differ considerably depending on the frequency
difference $|\omega_{pu} - \Omega_{i}|$. The difference is prominent in Rb as it has a narrower
Doppler width compared to Na.

         Using copropagating laser radiation the probe absorption always reveals three
strong dips corresponding to the hyperfine components and having the same separation as
the hyperfine splitting. There should be no crossover resonance as it is observed in the
case of Lamb dip spectroscopy. But it must be clear that the observed transmission peaks are not the hyperfine
transitions. The positions may be variable points of the
Doppler broadened curve depending solely on the pump laser frequency. It is found and
explained that the set of transmission peaks, that ''represent'' the hyperfine components,
tread along the Doppler profile and the pump laser actually pulls or drags the triplet
peaks. Thus we can control and manipulate the positions and intensities of transmission
peaks by shifting the position of the pump laser frequency. We can derive highly
accurate values of hyperfine splitting free from the problem of any crossover resonance
in a simple pump-probe experiment, though the absolute frequencies cannot be determined in this process.
No effect of Autler-Townes splitting can be found in this experiment.
The interesting feature of this experiment is that the triplet of 'hyperfine dips'
appear at all pump frequencies within the Doppler background.

Acknowledgments: P.N.G. thanks the BRNS, DAE,
New Delhi for the award of a research project. A.P and A.B thank the CSIR, New delhi for the award of Research Fellowships.
The project is also supported by the FIST programme of the DST, New Delhi.

\noindent \vskip 4.5mm

\noindent \vskip 4.5mm

\newpage \noindent {\bf References:}
\ben
\setlength{\itemsep}{0ex plus0.2ex}
\setlength{\parsep}{0.5ex plus0.2ex minus0.1ex}
\baselineskip 24pt

\item  M. Xiao, Y. Li, S. Jin and J. Banacloche, Phys. Rev. Lett. {\bf74}, 666(1995). 
\item  S. E. Harris, J. E. Field, A. Imamoglu, Phys. Rev. Lett. {\bf64}, 1107(1990).
\item  S. P. Tewari and G. S. Agarwal, Phys. Rev. Lett. {\bf56}, 1811(1986).  
\item  G. S. Agarwal and S. P. Tewari, Phys. Rev. lett. {\bf70}, 1417(1993).
\item  S. Wielandy and A. L. Gaeta, Phys. Rev. A {\bf58}, 2500(1998). 
\item  H. X. Chen, A. V. Durrant, J. P. Marangos and J. A. Vaccaro, Phys. Rev. A {\bf58}, 1545(1998). 
\item  Y. Zhu and T. N. Wasserlauf, Phys. Rev. A {\bf54}, 3653(1996).
\item  G. S. Agarwal, Phys. Rev. A {\bf55}, 2467(1997).
\item  R. R. Moseley, S. Shepherd, D. J. Fulton, B. D. Sinclair and M. H. Dunn, Phys. Rev. Lett. {\bf74}, 670(1995).
\item  V. S. Letokhov and V. P. Chebotayev, "Nonlinear Laser Spectroscopy". Springer-Verlag, Berlin (1977).
\item  W. Demtr$\ddot{o}$der, "Laser Spectroscopy". Springer-Verlag, Berlin (1982).
\item  H. R. Schl$\ddot{o}$sberg and A. Javan, Phys. Rev {\bf150}, 267(1966).
\item  S. Mandal and P. N. Ghosh, Phys. Rev. A {\bf45}, 4990(1992).
\item  C. E. Wieman, G. Flowers and S. Gilbert, Am. J. Physics. {\bf63}, 317(1995).
\item  D. Bhattacharyya, B. K. Dutta, B. Ray and P. N. Ghosh, Chem. Phys. Lett.(in-press).
\item  C. H. Townes and A. L. Schawlow, "Microwave Spectroscopy", Dover Publications, New York(1975).
\item  A. Javan, Phys. Rev. {\bf107}, 1579(1957).
\item  V. Wong, R. W. Boyd, C. R. Stroud, R. S. Vennink and A. N. Marino, Phys. Rev. A {\bf68}, 012502(2003).
\item  K. B. MacAdam, A. Steinbach, and C. Wieman, Am. J.Phys. {\bf60}, 1098(1992).
\een
\newpage 
\thispagestyle{empty}
\noindent{\bf $\:\,$Figure Captions} 
\newcounter{fig}
\begin{list}{\bf Figure \arabic{fig} :}{\usecounter{fig}
\setlength{\labelwidth}{2.cm} 
\setlength{\leftmargin}{1.37cm}
\setlength{\labelsep}{0.25cm} 
\setlength{\rightmargin}{0.1mm}
\setlength{\parsep}{0.5ex plus0.2ex minus0.1ex}
\setlength{\itemsep}{0ex plus0.2ex}} \baselineskip 21pt

\item Schematic representation of the experimental arrangement for 
      velocity-selective resonances; ECDL : external cavity diode laser,
      OI : optical isolator, BS : beam splitter, CBS : cubical beam splitter,
      PD : photo detector, M : mirror, L : convex lens, BD : beam dump.       

\item Energy level diagram for the double V configuration used in this work.
      F=2,3 for $^{85}$Rb and F=1,2 for $^{87}$Rb. $\Omega_i$ are resonance 
      frequencies for D$_2$ transition.

\item The probe absorption profile of $^{85}$Rb transition, when the pump laser
      is locked to different points of the Lamb dip spectrum of $^{85}$Rb
      5$^2S_\frac {1}{2}$ (F=3) $\rightarrow$ 5$^2P_\frac {3}{2}$ (F'=2,3,4):
      (a)the pump is switched off,
      (b)-(f) pump is on, the upward pointing arrows represent the lock points.
      
\item The probe absorption profile of $^{87}$Rb transition, when the pump laser
      is locked to different points of the Lamb dip spectrum of $^{87}$Rb
      5$^2S_\frac {1}{2}$ (F=2) $\rightarrow$ 5$^2P_\frac {3}{2}$ (F'=1,2,3): 
      (a) pump is off, (b)-(f) pump is on, the upward pointing arrows represent the pump lock points, while 
      the downward pointing arrows represent the velocity selective resonances.
\end{list}

\end{document}